\documentclass[12pt]{article}
\usepackage[pdftex]{color}
\usepackage{amsmath,amsthm,amssymb}
\usepackage{wasysym}
\usepackage{graphicx}
\usepackage{color}%
\usepackage[onehalfspacing]{setspace}
\usepackage{lipsum}
\usepackage[lmargin=3cm,rmargin=3cm,tmargin=3cm,bmargin=3cm]{geometry}
\usepackage{multirow}
\usepackage{hyperref}
\hypersetup{hidelinks}
\usepackage{mathtools}
\usepackage{breqn}
\usepackage{threeparttable}
\usepackage[utf8]{inputenc}
\usepackage{tocloft}
\usepackage{cite}
\usepackage{enumitem}

\setlist[itemize]{leftmargin=*}
\newcommand\keywords[1]{\textbf{Keywords}: #1}

\begin{document}
%\linenumbers
\begin{center}
{\rm \bf \Large{Non-monotonic evolution of contact area in soft contacts during incipient torsional loading }
}
\end{center}

\begin{center}
{\bf Bo Zhang$^{\text{ab}}$, Mariana de Souza$^{\text{c}}$, Daniel M. Mulvihill$^{\text{d}}$, Davy Dalmas$^{\text{c}}$\footnote{Corresponding author: davy.dalmas@ec-lyon.fr}, Julien Scheibert$^{\text{c}}$, Yang Xu$^{\text{ab}}$\footnote{Corresponding author: yang.xu@hfut.edu.cn}}
\end{center}

\begin{flushleft}
{
$^{\text{a}}$School of Mechanical Engineering, Hefei University of Technology, Hefei, 230009, China\\
$^{\text{b}}$Anhui Province Key Laboratory of Digital Design and Manufacturing, Hefei, 230009, China\\
$^{\text{c}}$CNRS, Ecole Centrale de Lyon, ENTPE, LTDS, UMR5513, 69134, Ecully, France\\
$^{\text{d}}$Materials and Manufacturing Research Group, James Watt School of Engineering, University of Glasgow, Glasgow, G12 8QQ, UK 
}
\end{flushleft}

\begin{abstract}
Many properties of soft contact interfaces are controlled by the contact area (e.g. friction, contact stiffness and surface charge generation). The contact area increases with the contact age at rest. In contrast, it usually reduces under unidirectional shear loading. Although the physical origin of such a reduction is still debated, it always happens in an anisotropic way because the reduction mainly occurs along the shearing direction. Whether such anisotropy is a necessary condition for shear-induced area reduction remains an open question. Here, we investigate the contact area evolution of elastomer-based sphere-plane contacts under an isotropic shear loading, i.e. torsional loading. We find that, when macroscopic sliding is reached, the contact area has undergone a net area reduction. However, the area evolves non-monotonically as the twisting angle increases, with an initial rise up to a maximum before dropping to the value during macroscopic sliding. The ratio of maximum to initial contact area is found weakly dependent on the normal load, angular velocity and dwell time (time interval between the instants when the normal load and twist motion are first applied) within the investigated ranges. We show that non-monotonic area evolution can also be found under unidirectional shear loading conditions under large normal force. These observations challenge the current descriptions of shear-induced contact area evolution and are expected to serve as a benchmark for future modelling attempts in the field. 
\end{abstract}

\keywords{Torsional loading; Unidirectional shear loading; Non-monotonic area evolution; Contact aging; Soft contact; PDMS}

\section{Introduction}
The morphology of the real contact between two solids (total area and its potentially complex distribution along the apparent contact) largely influences the main tribological properties of the contact interface including: static friction \cite{dieterich1994direct,li2011frictional,aymard2024designing}, sliding friction \cite{vorvolakos2003effects,krick2012optical,aymard2024designing}, surface charge generation \cite{min2021origin} and contact stiffness \cite{medina2013analytical}. Those properties are in turn crucially involved in many industrial and natural phenomena (e.g. high-precision positioning in extreme ultraviolet (EUV) lithography, triboelectric nanogenerator \cite{wu2019triboelectric}, and earthquakes \cite{scholz1998earthquakes}). At contact creation, the real contact is controlled by the normal load, curvatures and elastic moduli of the solids, and the surface roughness \cite{vakis2018modeling}. Even under constant normal load, the real contact then evolves as the result of an intricate competition between two main phenomena: aging (the increase of contact area with the time elapsed from initial contact) and variations of the contact area induced by the subsequently applied shear and sliding, giving rise, in particular, to the rate- and state-dependent friction law \cite{baumberger2006solid}. In soft materials like elastomers (typically PDMS/glass contacts), the contact area usually reduces during the incipient tangential loading of the contact \cite{savkoor1977effect, arvanitaki1995friction, waters2010mode,mergel2018continuum,sahli2018evolution, weber2019frictional,lengiewicz2020finite,peng2021effect}, by significant amounts of tens of percent. Note, however, that negligible reduction was reported in PDMS contact when the interface has low energy due to a specific coating \cite{vorvolakos2003effects, sahli2018evolution} and that sliding contact areas larger than the initial contact area were observed for rough nitrile rubber contacts \cite{krick2012optical}.

So far, shear-induced contact area reduction has been observed in experiments in which shear is applied along a specific direction (referred to as unidirectional shear loading below) \cite{savkoor1977effect, arvanitaki1995friction, waters2010mode,mergel2018continuum,sahli2018evolution, weber2019frictional,lengiewicz2020finite,peng2021effect,sahli2019shear,oliver2023adhesion}. In these conditions, reduction is observed to occur in an anisotropic way, because the contact area shrinks mainly along the shearing direction, so that initially circular contacts become ellipse-like \cite{waters2010mode,mergel2018continuum,sahli2018evolution,lengiewicz2020finite,sahli2019shear,oliver2023adhesion}. The micro-junctions in rough contacts also undergo anisotropic shrinking, mainly along the shear direction \cite{sahli2019shear}. Since significant shear-induced contact area reduction has always been observed simultaneously with the growth of a contact anisotropy, the question arises as to whether such anisotropy is a necessary condition for area reduction. In particular, understanding what happens to the contact area when the interface is subjected to shear applied in an isotropic way and not along a specific direction remains an open question. To answer this question, we have considered initially circular contacts, and subjected them to torsion around their symmetry axis (referred to as torsional loading below). In this way, the axisymmetry of the contact is not broken when shear is applied and one expects to avoid any anisotropy of the contact during its evolution during incipient torsional loading.

Torsional loading of soft sphere/plane contacts have previously been investigated. In particular, Chateauminois and co-authors measured the evolution of the in-plane displacement field and shear stress field along contact interfaces between glass spheres and flat thick PDMS layers, pressed under constant imposed normal displacement, when the twisting angle is increased \cite{chateauminois2010friction,trejo2013friction}. They documented the evolution of a partial slip regime with an inner stuck region of the contact that is shrinking upon increasing twist angle, while the peripheral slip region extends \cite{chateauminois2010friction}. For rough contacts, they also reported a constant shear stress in the slip region \cite{trejo2013friction}. Interestingly, both qualitative observations are identical to that made in similar contacts under unidirectional shear loading \cite{sahli2018evolution,lengiewicz2020finite,aymard2024designing}. Such similarities suggest that the elementary mechanisms involved in unidirectional shear loading and torsional loading are identical, and that only the symmetry of the way shear loading is applied is different. This confirms the interest of comparing both loading configurations to answer our question about the relationship between contact anisotropy and area reduction.

Below, we report on experiments that explore, during incipient torsional loading, the evolution of the contact between a soft spherical surface and a rigid flat using a customized opto-mechanical rig, with a special focus on the evolution of the contact area. We find that significant ($10-15\%$) area reduction occurs between initial contact and macroscopic sliding, and that this maximum reduction is reached after a non-monotonic evolution: the contact area first grows by typically $5\%$ and only then drops to its value during macroscopic sliding. We argue that this initial area growth is dominantly driven by shearing, rather than by aging.

The article is organized as follows: The experimental methods are provided in Section 2; the results in torsional loading are given in Section 3, complemented by those in unidirectional shear loading; a discussion on the possible mechanisms that underlie the observations is given in Section 4.

\section{Experimental methods}\label{sec:TCT}

\subsection{Torsional contact test}
\begin{figure}[h!]
  \centering
  \includegraphics[width=15cm]{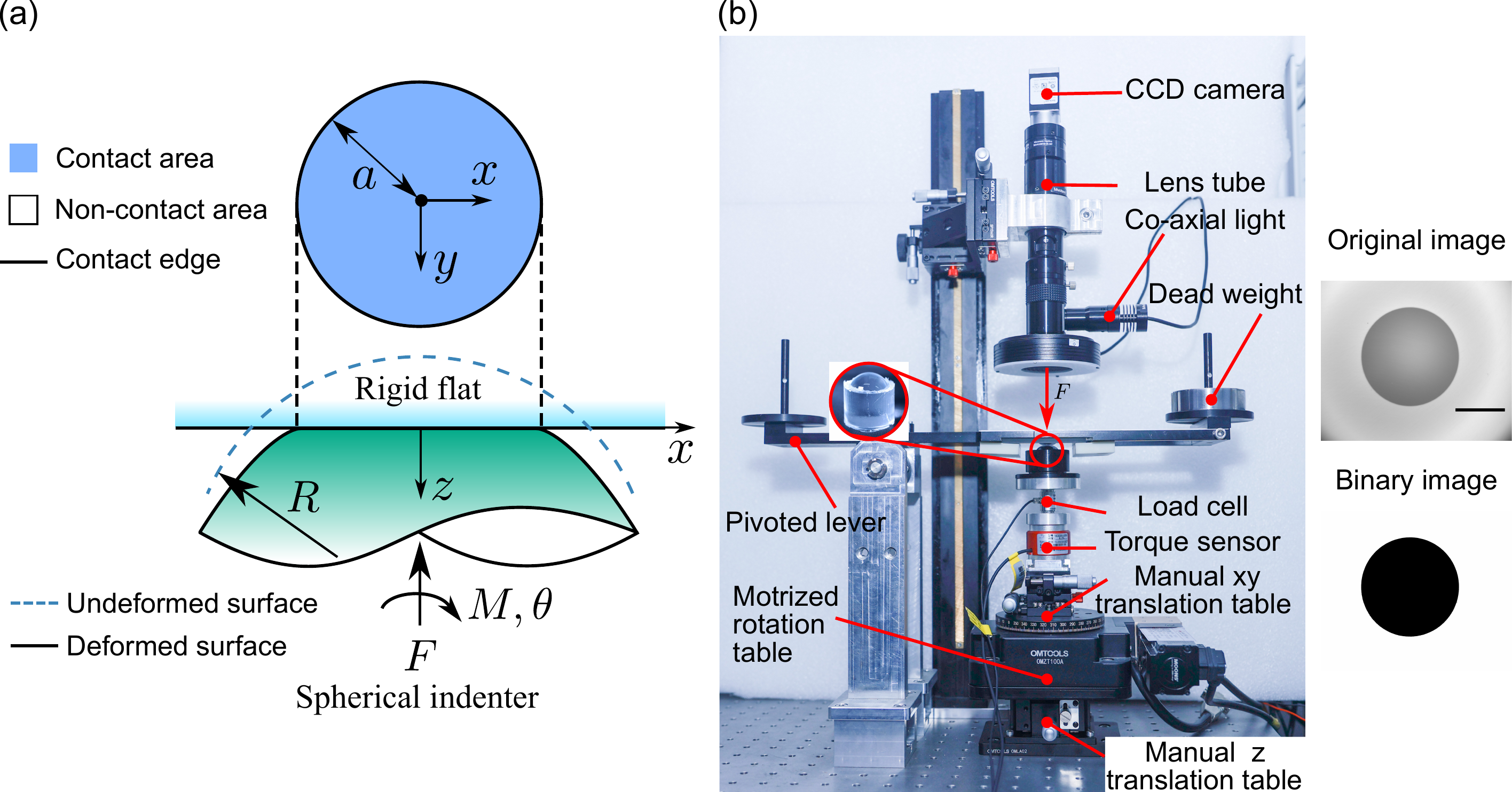}
  \caption{(a) Sketch of the torsional contact problem. Bottom: a soft spherical indenter of radius of curvature $R$ is submitted to a normal force $F$ and then to an increasing twist angle $\theta$, giving rise to a torque $M$. Top: sketch of the circular contact area of radius $a$. (b) Left: the opto-mechanical rig used in the experiments. Right: typical contact image (scale bar: $3$ mm) and corresponding segmented image, enabling measurement of the contact area.}\label{fig:Fig_1}
\end{figure}

An elastic spherical surface is in torsional contact with a rigid flat over a circular contact area of size $A = \pi a^2$ (see Fig. \ref{fig:Fig_1}(a)). The contact pair is subjected to a constant normal force $F$ and to an increasing twist angle $\theta$, which induces a resisting torque $M$. An opto-mechanical device has been built to apply and monitor such a torsional loading (see Fig. \ref{fig:Fig_1}(b)). The soft sphere of radius $R=13.11$ mm is made of PDMS (Sylgard 184, Dow, USA) with a base to curing agent ratio of 10:1. It has been fabricated by moulding \cite{delplanque2022solving} against an optical concave lens (GU OPTICS, China) and cured in the oven for $30$ minutes at a temperature of $120^{\circ}$C. The flat surface is made of a transparent N-BK7 soda-lime glass (Edmund Optics, USA). The root mean square roughness of the PDMS and glass plate surfaces, $248.8 \pm 110.5$ nm (for eight PDMS samples fabricated with the same procedure) and $138.2$ nm (the same glass sample is used for all tests) respectively, are negligible when compared to the maximum normal displacement of the PDMS surface during the tests. The normal force is applied on the contact pair through a pivoted lever. By adjusting deadweights on both sides of the lever, various normal forces can be achieved. A motorized rotation table (OMZT100A, OMTOOLS, China) drives the PDMS sphere in torsional contact against the glass plate with a constant angular velocity. A load cell ($20$ N, SBT641, SIMBATOUCH, China) and a torque sensor ($0.2$ Nm, SBT850A, SIMBATOUCH, China) are used to measure the normal force and torque acting on the PDMS sphere, respectively. 

As recommended by Chateauminois et al. \cite{chateauminois2010friction}, the offset between the rotation axis of the motorized table and the axis perpendicular to the contact area through the contact center was reduced, to a residual value of $80$ $\mu$m, by using a manual xy translation table (OMSDH02AE, OMTOOLS, China). Doing so, the offset-induced oscillation of the normal force during a complete rotation of the contact remains always smaller than $1\%$ of the mean normal force, for all experiments performed. It is considered negligible during the incipient torsional loading ($\theta \leq 100^{\circ}$). The contact area is optically monitored using a $10$ bits Charge-Coupled Device (CCD) camera (acA2440-20gm, Basler, Germany) with $5$ frames per second and $2448 \times 2048$ pixels, mounted on top of the glass plate. After automatically locating the periphery of the contact area in each recorded image, a corresponding binary image is obtained by setting the pixels inside (outside) the contact periphery to black (white) based on the assumption that complete contact occurs inside the contact periphery. This assumption is natural since both surfaces in contact are smooth and holds in almost all captured images in all experiments reported below (see Fig. \ref{fig:Fig_3}). A few exceptions contain some tiny isolated out-of-contact spots, which have a negligible effect on the measured contact area. The measured contact area is proportional to the number of black pixels. The size of a single pixel is calibrated using a negative 1951 USAF resolution test target (RTS3AB-N, LBTEK, China).

Before each test, the glass plate and the elastomer surface are cleaned thoroughly using alcohol and dried by blowing compressed air. The electrostatic charges on the glass and PDMS surfaces are effectively removed by using an ionizing fan (AP-DC2453, AP\&T, China). After application of the normal force, the torsional loading is initiated only after a controlled dwell time $t_{\text{d}}$ (also known as aging time). 

\subsection{Unidirectional shear test}

Unidirectional shear loading tests were conducted between a smooth PDMS sphere (Sylgard 184, Dow Corning, base to curing agent ratio $10:1$, cross-linked as described in Ref. \cite{lengiewicz2020finite}) with a radius of curvature $R=9.42$ mm and a rigid glass plate, using an opto-mechanical device already used in Ref. \cite{lengiewicz2020finite}. The PDMS sphere is first compressed against the glass plate under a constant normal load, during a dwell time smaller than $30$ seconds, before being sheared over a distance of $2$ mm at velocity $0.1$ mm/s. During shearing, the normal and tangential forces ($F$ and $Q$) are recorded and contact area images are taken at $100$ frames per second, enabling monitoring of the evolution of the contact area ($A$). The contact area is measured on a binary version of the raw grayscale images. Depending on the choice of segmentation threshold, the area, $A$, (resp. the contact sizes along and perpendicular to shear, $l_{\parallel}$ and $l_{\perp}$) may vary with a standard deviation representing $\pm 0.6\%$ (resp. $\pm 0.3\%$). Note that such an error on the absolute value of $A$, $l_{\parallel}$ and $l_{\perp}$ does not affect the subsequent relative evolutions of these quantities during shear (see Fig. \ref{fig:Fig_9} in Section \ref{subsec:ICG_friction}) and thus does not affect the main observation in those experiments (the non-monotonic trend of the contact area upon shear loading).

\section{Results}

\subsection{Torsional contact test}\label{subsec:torsion}
\begin{figure}[h!]
  \centering
  \includegraphics[width=16cm]{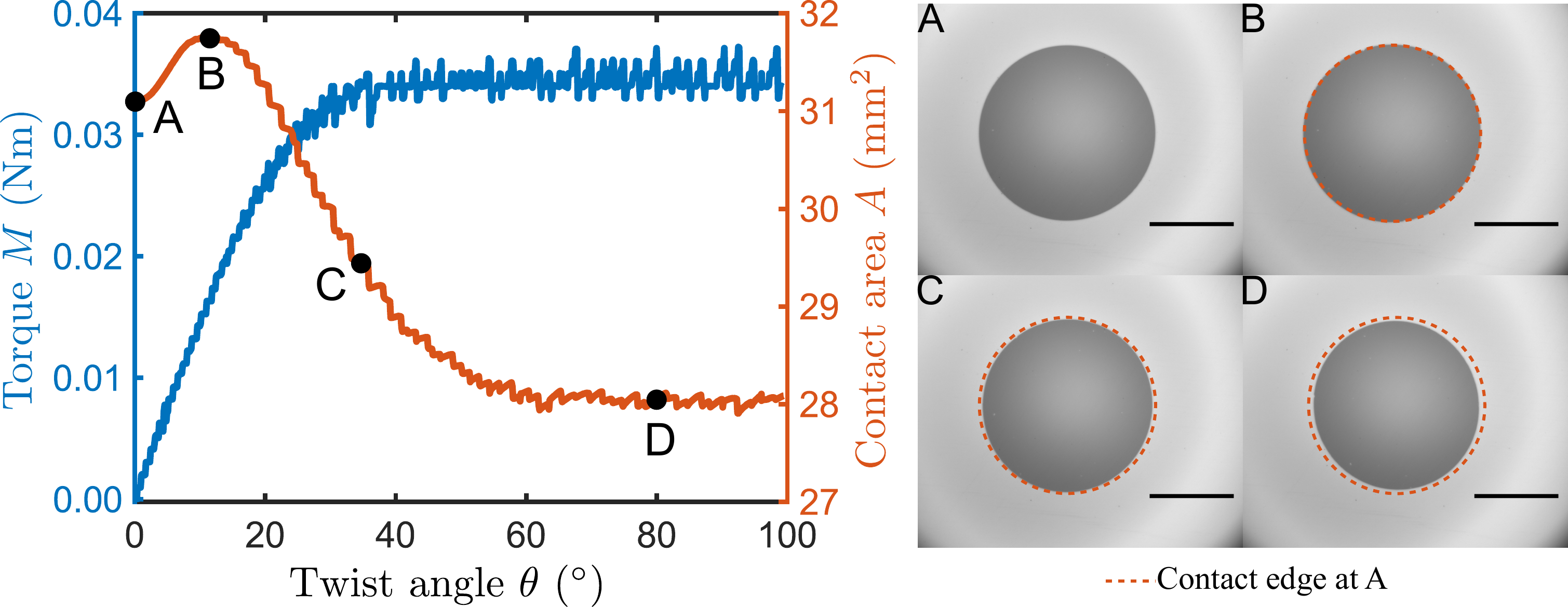}
  \caption{(Left) variations of torque and contact area with respect to twist angle during incipient torsional loading ($\theta<100^{\circ}$); (Right) contact images (scale bar: $3$ mm) taken at (A) $\theta=0^{\circ}$, (B) peak of $A(\theta)$, (C) $\theta=35^{\circ}$, (D) $\theta=80^{\circ}$. The mean normal force, angular velocity, and dwell time are $5.73$ N, $0.6^{\circ}$/s, and $30$ min, respectively.}\label{fig:Fig_3}
\end{figure}

Typical variations of the torque and contact area with twist angle during the incipient torsional loading are shown in Fig. \ref{fig:Fig_3}. In this particular example, the twisting motion initiates after the contact area has undergone aging during a $30$-min dwell time. The normal force remains approximately constant with negligible oscillation ($F = 5.73$ $\pm$ 0.02 N). The torque first linearly increases with twist angle, before weakening and smoothly transitioning toward a plateau when $\theta > 40^{\circ}$ (the curve exhibits small fluctuations due to the discontinuous rotation imposed by the stepper motor). Concurrently, the contact area is found to vary non-monotonically with the twist angle, i.e. it first smoothly grows by $2.1\%$ and then reduces by $11.3\%$ from the peak value with approximately the same rate. When the maximum contact area is reached, the torque is still half of its mean value during the final plateau. 

We emphasize that the combination of the $A(\theta)$ curve in Fig. \ref{fig:Fig_3} (left) and the observation that the contact area remains at all times circular during the torsional experiment (see images in Fig. \ref{fig:Fig_3}, right) already answers our initial question: contact anisotropy is not a necessary condition for contact area reduction. The surprising additional observation here is that the path to a reduced final area is non-monotonic, with an initial increase in area before the final drop.

\begin{figure}[h!]
  \centering
  \includegraphics[width=7cm]{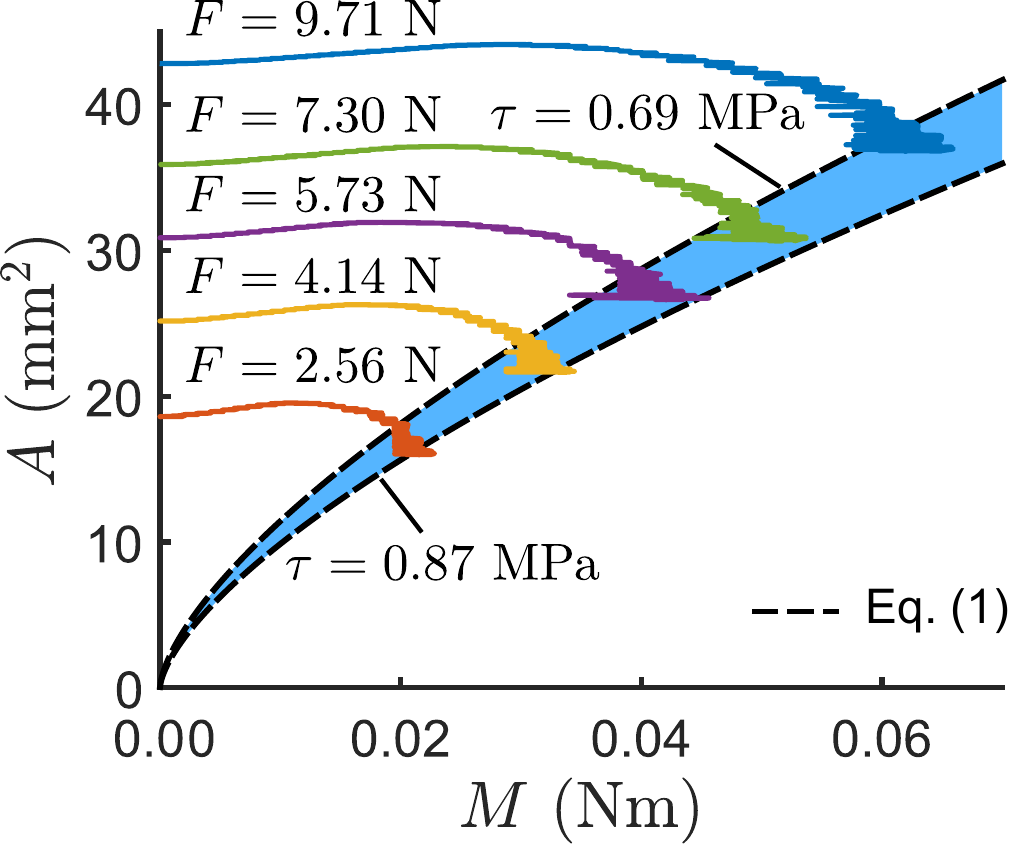}
  \caption{Variations of contact area with respect to torque under various normal forces (solid lines); The lateral oscillations at the right-side of all solid lines are due to oscillations of the torque during macroscopic sliding. Bottom/top dashed lines and blue shaded area correspond respectively to the plot of Eq. \eqref{eq:A_M_sliding} with $\tau = 0.87$ MPa, $\tau = 0.69$ MPa and $\tau \in (0.69, 0.87)$ MPa. Angular velocity and dwell time are $0.6^{\circ}/$s and $2$ min, respectively.}\label{fig:Fig_8}
\end{figure}

Figure \ref{fig:Fig_8} illustrates the contact area to torque relation, $A(M)$, under five different normal forces. As torque increases, the contact area first grows to a maximum value and then reduces till it terminates, sometimes with a significant lateral oscillation along the torque axis due to stick-slip-like torque oscillations. Assuming the interfacial shear stress is uniform and equal to the shear strength, $\tau$, the contact area to torque relation, when the interface is in its final, full sliding state, can be expressed by the following non-linear relation:  
\begin{equation}\label{eq:A_M_sliding}
A = \pi^{1/3} \left(\frac{3}{2 \tau}\right)^{2/3} M^{2/3}.
\end{equation}
Assuming that the lowest points of $M(\theta)$ in each torque fluctuation cycle in the plateau regime ($\theta>40^{\circ}$) are in full slip conditions (this assumption is presumably valid in stick-slip conditions), the shear strength of the interface can be estimated by using Eq. \eqref{eq:A_M_sliding}. Taking the mean value of the shear strength over all cycles within $\theta \in [70^{\circ}, 100^{\circ}]$, we found that the shear strength reduces with the increase of the normal force, from $\tau = 0.87$ MPa ($F = 2.56$ N) to $\tau = 0.69$ MPa ($F = 9.71$ N). This phenomenon of reduction of $\tau$ with $F$ is also detectable in friction tests with unidirectional shear loading (see Fig. 2C in Ref. \cite{sahli2018evolution}). The values of shear strength estimated in the current study are slightly larger than those reported for PDMS/glass contacts in the past literature ($\tau \leq 0.4$ MPa \cite{chaudhury2007studying}; $\tau \in [0.15, 0.20]$ MPa \cite{chateauminois2010friction}; $\tau = 0.36 \pm 0.01$ MPa \cite{sahli2018evolution}; $\tau = 0.43 \pm 0.01$ MPa \cite{mergel2018continuum}; $\tau = 0.281 \pm 0.001$ MPa \cite{oliver2023adhesion}; $\tau \in [0.225,0.450]$ MPa \cite{aymard2024designing}). This difference may be due to the different PDMS fabrication procedure and aging time adopted in the present study.  

\begin{figure}[h!]
  \centering
  \includegraphics[width=16cm]{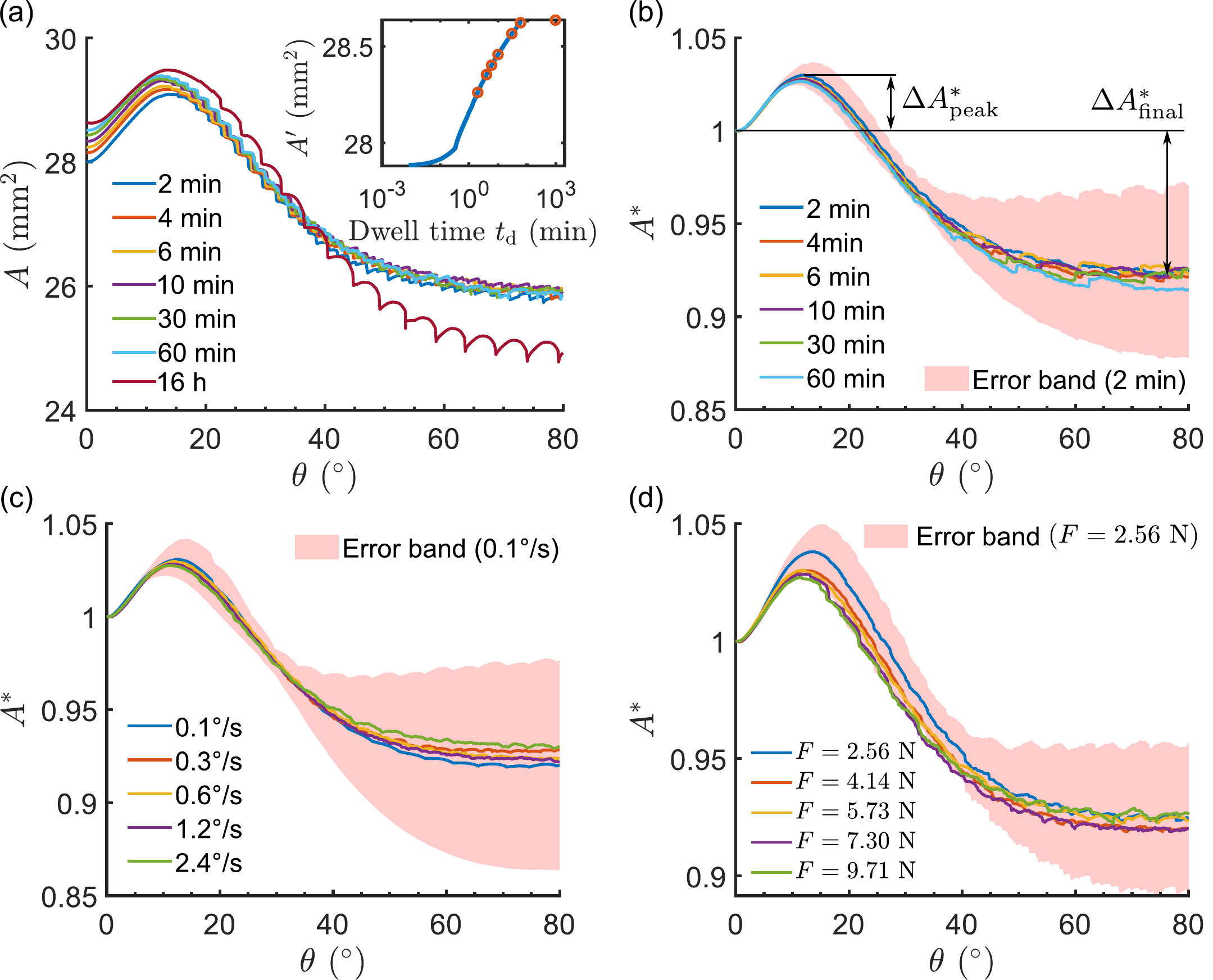}
  \caption{(a) Variation of $A$ with respect to twist angle for one typical PDMS sample during incipient torsional loading with different values of dwell time. The normal force and angular velocity are $5.73$ N and $0.6^{\circ}$/s, respectively; Inset: initial contact area $A'$ (i.e. for $\theta = 0$) as a function of dwell time (here for $F = 5.73$ N). Red hollow circular markers represent the values of contact area associated with the seven different values of dwell time used in the main panel ($2, 4, 6, 10, 30$, $60$ min, and $16$ h); (b) Variation of mean dimensionless $A^* = A(\theta)/A(\theta=0)$ with respect to twist angle. The normal force and angular velocity are $5.73$ N and $0.6^{\circ}$/s, respectively; (c) Variation of mean $A^*$ with respect to twist angle with various values of angular velocity. The normal force and dwell time are 5.73 N and $2$ min, respectively. (d) Variation of mean $A^*(\theta)$ with twist angle under various normal forces; Inset: variation of $A$ with twist angle under various normal forces. The angular velocity and dwell time are $0.6^{\circ}$/\text{s} and $2$ min, respectively. In panels (b-d), pink regions represent error bands over the eight tested PDMS samples.}\label{fig:Fig_4567}
\end{figure}

We investigated three factors that possibly influence the non-monotonic variation of area during incipient torsional loading, namely the dwell time, angular velocity, and normal force. The torsional contact tests (except for the 16-hour dwell time test) have been repeated eight times using different PDMS samples following the same fabrication procedure and the same glass plate. Typical results associated with one PDMS sample and the mean results with error bands for the eight PDMS samples are shown in Fig. \ref{fig:Fig_4567}(a) and Fig. \ref{fig:Fig_4567}(b-d), respectively. Presumably due to viscoelastic creep deformation of the soft bulk material, the contact area grows with dwell time when the PDMS surface is in purely normal contact with a glass plate before any torsion is applied \cite{Ovcharenko08,weber2019frictional}. The inset of Fig. \ref{fig:Fig_4567}(a) illustrates, only for the constant normal force ($F = 5.73$ N), such growth of the contact area (denoted as $A'$) with dwell time. The initial growth of the contact area is rapid, and the growth rate quickly drops after $10$ min. This growth trend is similar to that reported by Ovcharenko et al. \cite{Ovcharenko08}. Note that the logarithmic growth of the contact area, which has been found, e.g. by Weber et al. \cite{weber2019frictional} for a rough spherical contact, is not followed in this study. At the end of the dwell time, the twisting motion starts. To characterize the relative changes of contact area due to torsion, we define $A^*=A(\theta)/A(\theta=0)$, i.e. the ratio between the area of the twisted contact with twisting angle $\theta$ and that when $\theta=0$ (but after the dwell time). In particular, we define $\Delta A^*_{\text{peak}}$ (resp. $\Delta A^*_{\text{final}}$) as the absolute value of the difference between $A^*$ at $\theta=0$ and at the peak (resp. in the final plateau $\theta > 80^{\circ}$), see the schematic in Fig. \ref{fig:Fig_4567}(b).

Figure \ref{fig:Fig_4567}(a) shows the $A(\theta)$ curves of a typical PDMS sample for seven different values of dwell time. It should be noted that $A(\theta = 0)$ grows monotonically with dwell time, which is consistent with the inset of Fig. \ref{fig:Fig_4567}(a). For dwell time ranging from $2$ min to $60$ min, the dimensional evolution of contact area, $A(\theta)$, gradually converges to a master curve during area reduction, which ends up with a dwell time-independent $\Delta A_{\text{final}} = \Delta A_{\text{final}}^* \times A(\theta = 0)$. This implies that the effect of creep deformation on the contact area gradually becomes negligible as $A$ reduces toward a relatively constant value once macroscopic sliding is reached. The evolution of $A(\theta)$ with the dwell time observed for a typical PDMS sample is also followed qualitatively by mean $A(\theta)$ for the eight samples. The dimensional $A(\theta)$ with 16-hour dwell time gradually deviates from those with $t_{\text{d}} \leq 60$ min and ends up with a dimensional $A_{\text{final}}$ significantly lower than those shorter dwell time results. This difference is presumably not caused by the creep deformation, which can only result in contact area growth and seems to saturate after $t_{\text{d}} > 60$ min (see inset of Fig. \ref{fig:Fig_4567}(a)).

The same tests in Fig. \ref{fig:Fig_4567}(a) are repeated for all eight samples, and the evolution of the dimensionless mean $A^*(\theta)$ with respect to dwell time are shown in Fig. \ref{fig:Fig_4567}(b). All curves are identical initially, but start to bifurcate at $\theta \approx 10^{\circ}$. Beyond the bifurcation point, the overall $A^*(\theta)$ reduces as dwell time increases. Figure \ref{fig:Fig_4567}(b) shows that $\Delta A^*_{\text{peak}}$ slightly increases with the decreasing dwell time, while $\Delta A^*_{\text{final}}$ becomes larger at higher dwell time. Similar non-monotonic variation can be found in Fig. \ref{fig:Fig_4567}(c) for all investigated angular velocity ranging from $0.1^{\circ}$/s to $2.4^{\circ}$/s. The beginning of the mean curve $A^*(\theta)$ is nearly independent of the investigated angular velocity. Figure \ref{fig:Fig_4567}(c) indicates that $\Delta A^*_{\text{peak}}$ and $\Delta A^*_{\text{final}}$ respectively decreases and increases with increasing angular velocity. At small angular velocities, a stick-slip-like motion is visible for some samples, which is responsible for large oscillations of the contact area (also seen in Fig. \ref{fig:Fig_4567}(a) for $t_{\text{d}}=16$ h). Similar oscillating contact area has recently been observed in friction tests under unidirectional shear loading \cite{lengiewicz2020finite,lyashenko2024transition}. Note that we observed that torsional contacts with angular velocities larger than $2.4^{\circ}/\text{s}$ result in surface wrinkling and an unstable pattern of contact area \cite{chaudhury2007studying,yashima2019shearing}, thus, these are not shown here. The mean contact area evolution strongly depends on the normal force (see the inset of Fig. \ref{fig:Fig_4567}(d)). The main panel of Fig. \ref{fig:Fig_4567}(d) shows that $\Delta A^*_{\text{peak}}$ decreases with increasing normal force. In contrast, $\Delta A^*_{\text{final}}$ is found rather constant with the normal force. The latter behaviour is different from that found in Ref. \cite{sahli2018evolution}, where the relative decrease of area reached in full sliding was found decreasing when the normal force was increased.

The error bands (standard deviation of $A^*(\theta)$ for the eight samples) of curves in the same result group (i.e. Fig. \ref{fig:Fig_4567}(b, c, d)) are approximately the same. The dispersion of $A^*$ for a fixed twist angle is negligible before contact area reaches its peak value (mean standard deviation for all investigated cases within $\theta \leq 10^{\circ}$ is 0.0018), and increases by more than 10 times afterward (mean standard deviation for all investigated cases within $\theta > 10^{\circ}$ is $0.021$). The abrupt increase of the standard deviation is mainly due to randomly occurred stick-slip-like area oscillations for some samples. Changes of $A^*(\theta)$ caused by various influence factors are comparable or even significantly less than the standard deviation beyond the bifurcation point. Thus, the non-monotonic evolution of $A^*(\theta)$ during incipient torsional loading is weakly dependent on the dwell time, angular velocity and normal forces within the investigated ranges. Overall, the thinness of the error band at the beginning of the curves demonstrate the statistical robustness of the initial non-monotonic behaviour.

In summary, the contact area varies non-monotonically during incipient torsional loading in all investigated cases with different values of dwell time, angular velocity, and normal force. $\Delta A^*_{\text{peak}}$ is always found in the $3$-$5\%$ range, while $\Delta A^*_{\text{final}}$ can vary more, within the $4$-$15\%$ range.

\subsection{Unidirectional shear test}\label{subsec:ICG_friction}

The non-monotonic variation of contact area shown in Section \ref{subsec:torsion} is actually not specific to torsional loading. Indeed, we found similar non-monotonic trends during friction tests in unidirectional shear loading conditions, provided they are performed at relatively high normal forces. 

\begin{figure}[h!]
  \centering
  \includegraphics[width=16cm]{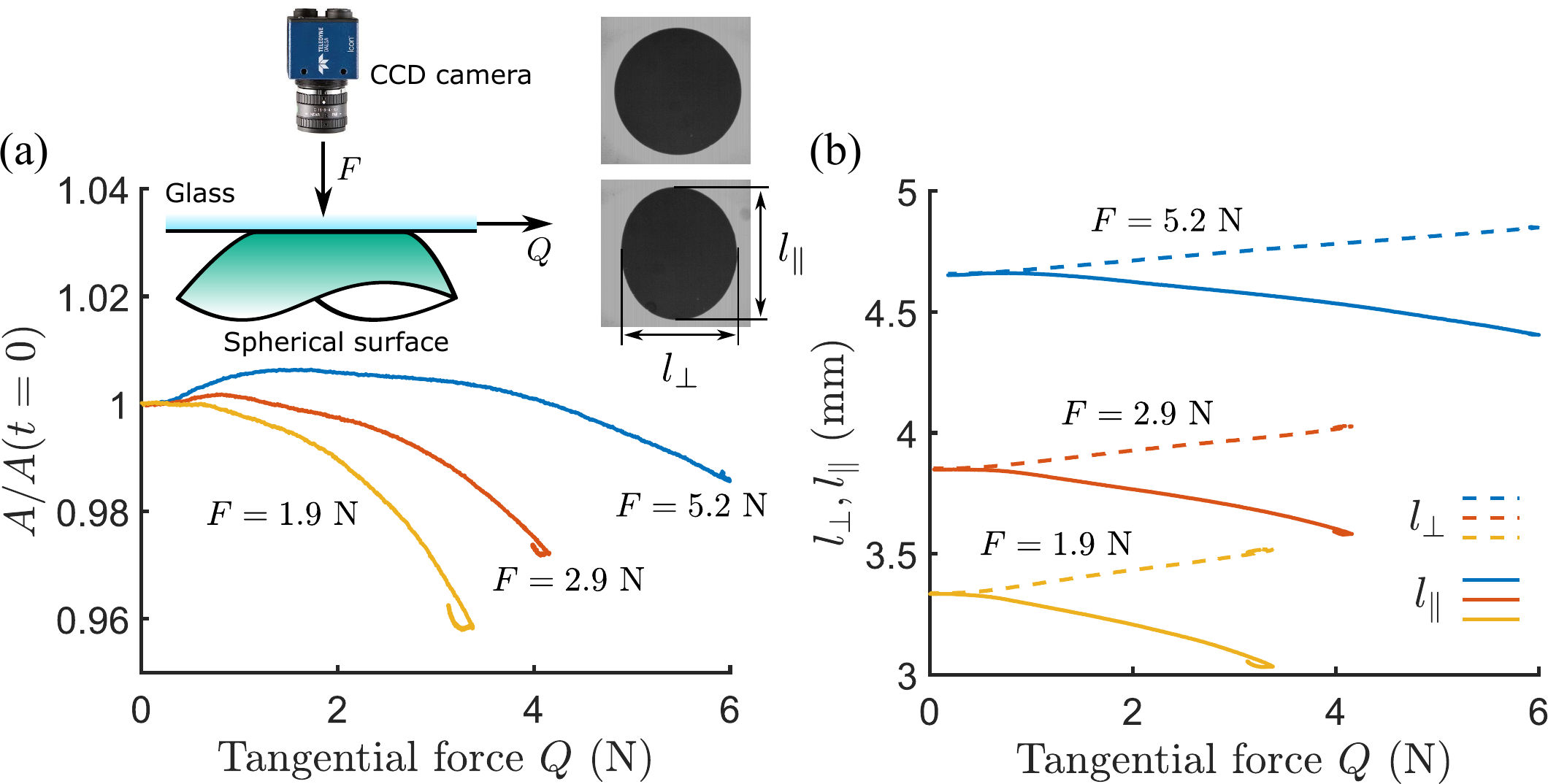}
  \caption{
  (a) Contact area $A(t)$, normalized by its initial value at the beginning of the test $A(t=0)$, as a function of the tangential force $Q$, for a unidirectionally sheared contact between a PDMS sphere and a glass plate, under three different normal forces ($F = 1.9$ N, $2.9$ N, and $5.2$ N). Left inset: sketch of the test; Right inset: two contact images at $F = 1.9$ N, before shearing (top, contact diameter is $3.34$ mm) and during full sliding (bottom). The PDMS sphere moves relative to the glass plate toward the left. (b) Evolution of the contact sizes parallel and perpendicular to the shear direction ($l_{\parallel}$ and $l_{\perp}$, respectively, see inset of (a)) as functions of the tangential force $Q$, for the same three experiments as in (a).}\label{fig:Fig_9}
\end{figure}

Figure \ref{fig:Fig_9}(a) shows the evolution of the relative change of the contact area with the tangential force. The contact area first increases (or remains approximately constant for the smallest normal force $F=1.9$ N), followed by a reduction, similar to what has been commonly observed in the literature \cite{savkoor1977effect,waters2010mode,mergel2018continuum,sahli2018evolution,weber2019frictional,lengiewicz2020finite,sahli2019shear}. During the incipient loading, the initially circular contact area progressively evolves to an ellipse-like anisotropic area (see the inset of Fig. \ref{fig:Fig_9}(a)). The relative growth of the contact area in the frictional contact, about $1\%$ for $F=5.2$ N, is found smaller than that in the torsional contact. Note that the qualitative behaviours seen in Fig. \ref{fig:Fig_9} are typical: an initial increase of contact area has indeed been found in more than half of $48$ similar experiments performed on $18$ different PDMS samples fabricated with the same procedure, and using normal loads larger than 2 N (up to $5.4$ N). The growth of the shear-induced anisotropy is further characterized through the variation of the two contact sizes ($l_{\parallel}$ and $l_{\perp}$) of the ellipse-like contact area, as plotted in Fig. \ref{fig:Fig_9}(b). Therein, $l_{\parallel}$ decreases while $l_{\perp}$ increases as the tangential force increases. The initial area increase is due to an increase of $l_{\perp}$ which is faster than the decrease of $l_{\parallel}$.

Such overall contact area reduction in unidirectional shear loading conditions has been attributed in the modelling literature to a (not yet fully uncovered) combination of adhesion \cite{peng2021effect,papangelo2019mixed,salehani2019modeling} and large deformation of the PDMS substrate \cite{lengiewicz2020finite}. In contrast, the non-monotonic evolution of the contact area has not been reproduced by any model yet. Experimentally, we interpret the fact that it has not been reported before to the large normal forces (up to $5$ N) used here, compared to the smaller forces (below $2.2$ N) used in previously published experiments with the same PDMS and sphere radius \cite{mergel2018continuum,sahli2018evolution,lengiewicz2020finite,sahli2019shear,guibert2021versatile,oliver2023adhesion}. Figure \ref{fig:Fig_9} thus demonstrates that non-monotonic shear-induced evolutions of the contact area can occur in both unidirectional shear and torsional loading configurations. Since the shearing of a contact interface is generally a superposition of a translational and a torsional motion, the observed non-monotonic trend may serve as a general phenomenological description of the evolution of the contact area during the incipient loading.

\section{Discussion}\label{sec:Discussion}
We now discuss the possible origin of the non-monotonic evolution of the contact area during incipient torsional loading. The existence of a peak suggests a competition between an increasing and decreasing trend, the former and the latter being dominant in the early stage and the final stage of the loading, respectively.

For the decreasing trend that dominates under large shear, it is natural to hypothesize that the same phenomena are at play in both the torsional and unidirectional shear configurations. For the latter, an abundant literature exists, where two main mechanisms have been proposed. Many linear elastic models have interpreted the shear-induced contact area reduction as a weakening of the effective work of adhesion when the contact is submitted to a combination of normal and tangential loading \cite{savkoor1977effect,waters2010mode,peng2021effect,oliver2023adhesion,papangelo2019mixed,papangelo2019shear,mcmeeking2020interaction,xu2022asperity}. However, in such linear elastic contacts, when the frictional shear stress is approximately uniform \cite{chateauminois2010friction}, as is likely in our experiments, no adhesion weakening effect is expected \cite{menga2018uniform,menga2019corrigendum}. And indeed, contact area reduction in elastomers can be captured quantitatively by an adhesionless elastic model that accounts for the Tresca friction model and large deformations of the bulk of the sphere \cite{lengiewicz2020finite}, which are expected from the large ratio of shear strength to shear modulus (e.g. above $50\%$ in \cite{sahli2018evolution,lengiewicz2020finite,sahli2019shear}). In this context, models incorporating both adhesion and finite strain (large deformation) have suggested that adhesion, although not at the origin of contact area reduction, may have an amplifying effect on it \cite{mergel2021contact}. In conclusion, for torsional loading, we expect that the same mechanisms (finite strain possibly amplified by adhesion) are relevant to explain qualitatively the observed overall area reduction. However, their quantitative combination is likely different, due to the different strain fields involved in either torsional or unidirectional shear loading.

It is interesting to examine in more detail how the contact morphology evolves anisotropically under unidirectional shear loading. Figure \ref{fig:Fig_9}(b) shows that, while $l_{\parallel}$ decreases under shear, $l_{\perp}$ increases concurrently. This observation indicates that the contact periphery moves differently at different locations relative to the shear direction: (i) when shear is orthogonal to the contact periphery (leftmost and rightmost contact points in the images of Fig. \ref{fig:Fig_9}(a)), the contact periphery moves inward, so that $l_{\parallel}$ decreases; (ii) when shear is parallel to the contact periphery (uppermost and bottommost points in the images of Fig. \ref{fig:Fig_9}(a)), the contact periphery moves outward, so that $l_{\perp}$ increases. Assuming that the same phenomenology would hold in torsional loading, and observing that the circular contact periphery is at all points parallel to the local shear direction, one would expect the periphery to move outward everywhere. Such a conclusion would be consistent with the observations of both a global increase of area and of a larger initial increase under torsional loading (the periphery moves outward at all points) than under unidirectional shear loading (the periphery moves outward only over a portion of the periphery). Testing the validity of such a scenario will likely require numerical simulations similar to those of Refs. \cite{lengiewicz2020finite,mergel2021contact}.

\begin{figure}[h!]
  \centering
  \includegraphics[width=6cm]{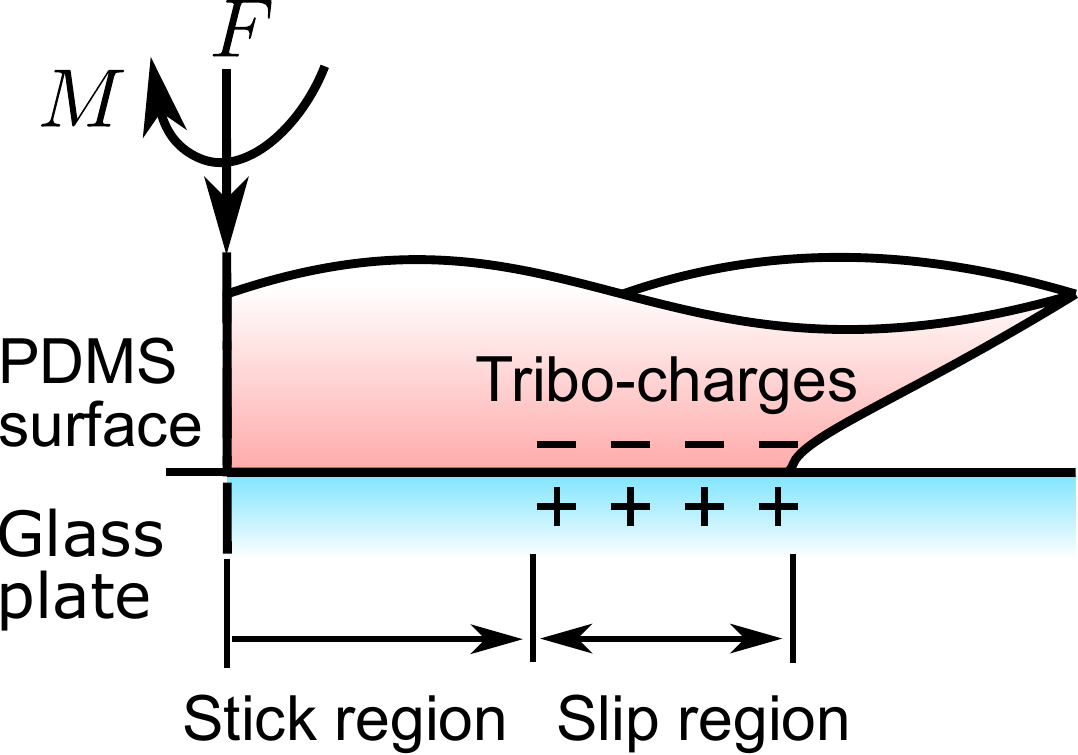}
  \caption{Schematic of a charged torsional contact interface where tribo-charges are assumed to transfer from glass plate to PDMS surface.}\label{fig:Fig_12}
\end{figure}

Let us now examine other mechanisms expected to produce an increasing trend for the contact area. First, one can imagine an adhesion strengthening effect from the tribo-charges transferred across the frictional interface as slip develops within the contact \cite{budakian2000correlation}. As the twist angle increases from zero, a slip zone initiates from the contact area periphery and monotonically expands inward toward the contact center \cite{chateauminois2010friction}. Electrostatic charges may accumulate over the annular slip zone with an increasing total charge. Thus, the interface would be similar to a charged capacitor (see Fig. \ref{fig:Fig_12}), and the effective work of adhesion may increase. Testing such a scenario would require measuring surface charges using either a Faraday cup, proof plane or Kelvin probe force microscope in an ideal environment \cite{lowell1980contact}, a task that is beyond the scope of the present study.

\begin{figure}[h!]
  \centering
  \includegraphics[width=8cm]{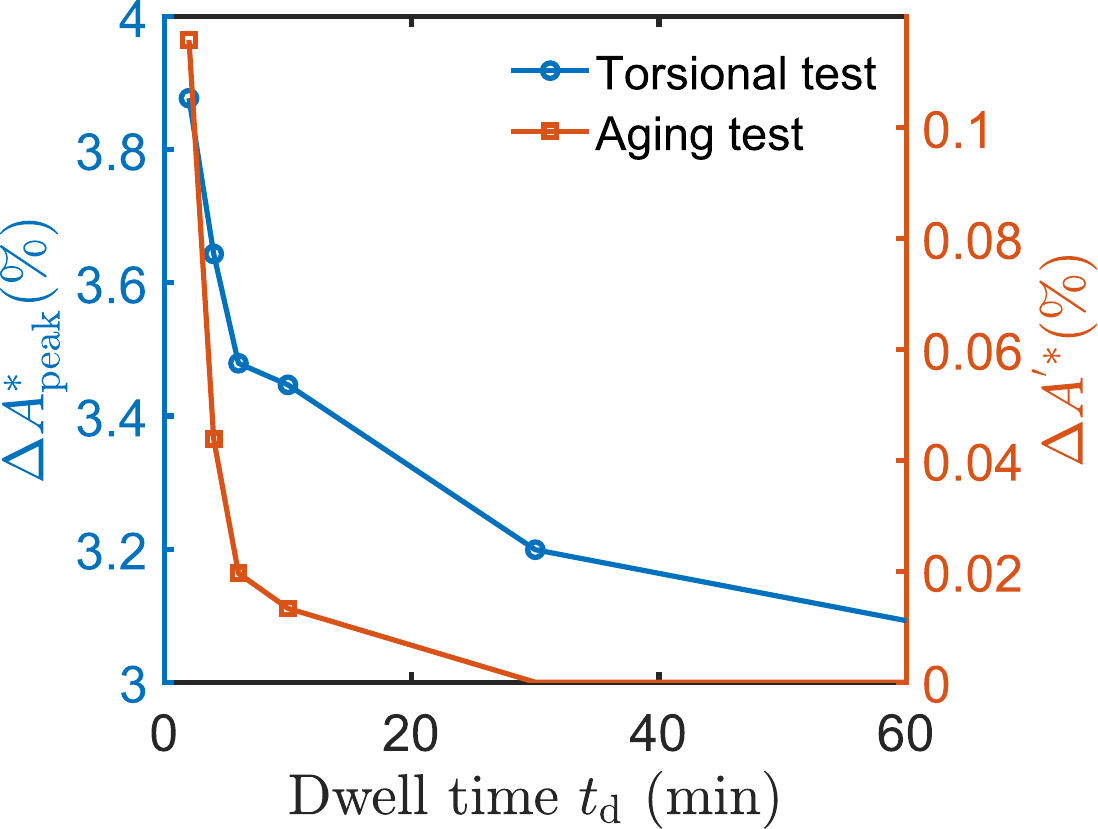}
  \caption{Variations of $\Delta A^*_{\text{peak}}$ and $\Delta A^{'*}$ with dwell time in the torsional and aging tests, respectively. The normal force and angular velocity are $5.73$ N and $0.6^{\circ}$/s, respectively.}\label{fig:Fig_11}
\end{figure}

A natural alternative mechanism for contact area increase is that underlying the contact growth during aging at rest (see inset of Fig. \ref{fig:Fig_4567}(a)). In elastomers, this mechanism is presumably viscoelastic creep \cite{ronsin2001state}. The decelerating rate of aging suggests that aging-induced area growth may dominate the finite strain/adhesion weakening-based area reduction at the beginning of the shear loading only, consistent with an initial net increase of the area, a maximum and then a final decrease. To test this hypothetical scenario, we compare the relative increase of area observed at the peak of $A(\theta)$ ($\Delta A^*_{\text{peak}}$, see Fig. \ref{fig:Fig_4567}(b) and Section 3) and the expected relative increase of area due to aging alone, taken over the same time interval $[t_{\text{d}} , t_{\text{peak}}]$, i.e. during the period from the end of the dwell time/start of shearing to the peak in contact area ($\Delta A^{’*}=(A’(t=t_{\text{peak}})- A’(t=t_{\text{d}}))/ A’(t=t_{\text{d}})$, where $A’(t)$ is the aging evolution of the contact area under purely normal load as a function of the contact time $t$, see inset of Fig. \ref{fig:Fig_4567}(a)). $A’(t=t_{\text{peak}})$ and $A’(t=t_{\text{d}})$ have been calculated from interpolations of the curve in the inset of Fig. \ref{fig:Fig_4567}(a). Both quantities are plotted in Fig. \ref{fig:Fig_11}, which clearly shows that $\Delta A^*_{\text{peak}} \gg \Delta A^{’*}$ for all investigated values of the dwell time $t_{\text{d}}$. The magnitude of the time interval needed to reach the peak, $t_{\text{d}}- t_{\text{peak}}$, is significantly smaller than $t_{\text{d}}$ (for instance, $t_{\text{d}} - t_{\text{peak}} \approx 23$ sec for $0.6^{\circ}/$sec, which is about one order of magnitude less than the minimum investigated dwell time ($2$ min)). Thus, the increase of contact area due to aging of the contact under purely normal load cannot be the dominant reason for the initial area increase during torsional loading. In this context, the observed increase is intrinsically an effect of shear. Note that similar shear-amplified increase of contact area has already been observed in Ref. \cite{weber2019frictional} in glassy polymer rough contacts, although the physical origin may be different (plasticity-related rather than viscoelasticity-related aging).

\section{Conclusion}
From the study of the smooth torsional contact between a PDMS sphere and a glass plate, we have shown that contact anisotropy is not a necessary condition for shear-induced contact area reduction. We have discovered that, in this loading mode, the contact area varies non-monotonically: it first increases and then decreases as the twist angle/torque increases during the incipient torsional loading. This phenomenology has been observed in all explored conditions, with various dwell times, angular velocities and normal forces. We have also managed to qualitatively observe similar non-monotonic area evolution under unidirectional shear conditions, by using larger normal forces than those reported previously in the literature.

The precise origin of the non-monotonic shear-induced area evolution remains to be fully unravelled. Nevertheless, we have argued that the mechanisms usually invoked for unidirectional tests (finite strain and adhesion) are likely also relevant in torsional tests. We have also shown that classical contact aging (area growth with time under constant purely normal force) is not responsible for the observed initial increase of area, which is thus intrinsically an effect of shear.

Overall, our results suggest that non-monotonic contact area evolutions is a more general phenomenological description of the incipient loading of sheared contact than the picture of a mere reduction that prevails in the literature. It may be particularly useful when studying interacting surfaces subjected to more complex tangential loading (a combination of translational and torsional loading). Our measurements challenge the existing models for sheared contact as, to our best knowledge, none of them has predicted such non-monotonic area evolutions. Future studies should aim at identifying the interplay of elementary mechanisms responsible for the observed non-monotonic behaviour, which will likely involve further experiments and enriched modelling.

\section*{Acknowledgements}
This work was supported by the National Natural Science Foundation of China (52105179) and Fundamental Research Funds for the Central Universities (JZ2023HGTB0252, PA2024GDSK0044). This work was carried out as part of the Pregliss project supported by the Carnot institute Ingénierie@Lyon, labeled by the French National Research Agency. YX and BZ would like to thank Prof. Ying Hu (Hefei University of Technology) for providing facilities to fabricate PDMS samples and Mr. Yixin Chen for taking the photograph of the torsional contact rig in Fig. \ref{fig:Fig_1}(b). 

\section*{Data availability statement}
The data that support the findings of this study are available from the corresponding authors upon reasonable request.

\begin{spacing}{1} % 行距
	\bibliographystyle{asmejour}
	\bibliography{ref}
%	\vspace{11bp}
\end{spacing}
\end{document}